\documentclass[final]{aipproc}

\layoutstyle{6x9}
\usepackage[english]{babel}
\usepackage{amsmath}
\usepackage{amssymb}
\usepackage{epsfig}
\usepackage{graphics,psfrag,rotating}
\usepackage{graphicx}

\begin{document}
\newcommand{\be}{\begin{equation}}
\newcommand{\ee}{\end{equation}}
\newcommand{\bea}{\begin{eqnarray}}
\newcommand{\eea}{\end{eqnarray}}
\newcommand{\beaa}{\begin{eqnarray*}}
\newcommand{\eeaa}{\end{eqnarray*}}
\newcommand{\Lhat}{\widehat{\mathcal{L}}}
\newcommand{\nn}{\nonumber \\}
\newcommand{\e}{\mathrm{e}}

\title{Conformal frames and the validity of Birkhoff's theorem}


\author{S. Capozziello}{
  address={Dipartimento di Scienze Fisiche, Universit\`a di Napoli ``Federico II'' and 
INFN Sez. di Napoli, Compl. Univ. Monte S. Angelo, Ed.N, Via Cinthia, I-80126 Napoli, Italy, EU}
}
\author{D. S\'aez-G\'omez}{
  address={Institut de Ci\`encies de l'Espai (ICE-CSIC/IEEC),
Campus UAB, Facultat de Ciencies, Torre C5-Par-2a pl, E-08193
Bellaterra (Barcelona), Spain, EU}}

\begin{abstract}
Birkhoff's theorem is one of the most important statements of Einstein's general relativity, which generally can not be extended to modified theories of gravity. Here we study the validity of the theorem in scalar-tensor theories using a perturbative approach, and compare the results in the so-called Einstein and Jordan frames. The implications of the results question the physical equivalence between both frames, at least in perturbations.
\end{abstract}

\maketitle
%

\section{Introduction}
The aim of this work is the study of spherically symmetric solutions  and the validity of  Birkhoff's theorem for non-minimally coupling scalar-tensor theories, and its relation with its conformal theory in the so-called  Einstein frame (for a review on higher order theories of gravity and conformal transformations, see Ref.~\cite{Capozziello:2011et}). The mathematical equivalence between both frames is a well known fact, while its physical relation is still an old and open discussion in classical gravitation ( see Ref.~\cite{Faraoni:1998qx}). Here, we are interested to compare both frames by the analysis of spherically symmetric solutions, and discuss the validity of Birkhoff's theorem using a perturbative approach (for more details see Ref.~\cite{Capozziello:2011wg}). Let us start by writing  the general action for a non-minimally scalar-tensor theory,
\be
S_{BD}=\int d^4x\sqrt{-g}\left[\phi R-\frac{\omega}{\phi}g^{\mu\nu}\nabla_{\mu}\phi\nabla_{\nu}\phi-V(\phi)+2\kappa^2\mathcal{L}_m\right]\ .
\label{1.4}
\ee
Here we assume $\omega$ to be a constant. Note that the action (\ref{1.4}) is equivalent to $f(R)$ gravity when the kinetic term is null, $\omega=0$, Ref.~\cite{reconstruction1}. 
This  action is commonly said to be expressed in the Jordan frame, while  the Einstein frame is recovered by applying a conformal transformation,
 \be
g_{E\mu\nu}=\Pi^2g_{ \mu\nu}, \quad \rightarrow \quad S_{E}=\int d^4x\sqrt{-g_E}\left[R_E-\frac{1}{2}\nabla_{\mu}\varphi\nabla^{\mu}\varphi-U(\varphi) +2\kappa^2\mathcal{L}_{Em}\right]\ .
\label{2.1}
\ee
where $\Pi^2=\phi$. Here the subscript $_E$ denotes  the variables  defined in the Einstein frame. For convenience, we have redefined the scalar field  as $\phi=\e^{\varphi/\sqrt{3+2\omega}}$, while the scalar potential is given by $U(\varphi)=\e^{\varphi/\sqrt{3+2w}}V(\phi(\varphi))$, and the matter Lagrangian is $\mathcal{L}_{Em}=\frac{1}{\phi^2}\mathcal{L}_m\left(\frac{1}{\phi^2}g_{E\mu\nu}\right)$.  The field equations can be  obtained by varying the action (\ref{2.1}) with respect to  $g_{E \mu\nu}$ and $\phi$,
\be
 R_{E\mu\nu}-\frac{1}{2}g_{E\mu\nu}R_E=\frac{1}{2}\partial_{\mu}\varphi \partial_{\nu}\varphi-\frac{1}{2}g_{E\mu\nu}\left[\partial_{\sigma}\varphi \partial^{\sigma}\varphi+U(\varphi)\right]+\kappa^2T^{(m)}_{E\mu\nu}\ ,
\label{2.4}
\ee
\be
\Box\varphi-\frac{dU(\varphi)}{d\varphi}=-2\kappa^2\frac{\delta(\mathcal{L}_{Em})}{\delta\varphi}\ ,
\label{2.5}
\ee
where the energy-momentum tensor is given by $T_{E\mu\nu}^{(m)}=\frac{-2}{\sqrt{-g_E}}\frac{\delta\mathcal{L}_{E\mathrm{m}}}{\delta g_E^{\mu\nu}}$.  We are interested to study perturbations around a given background solution, and find the range of validity of the Birkhoff's theorem for both frames. Particularly we assume in the Jordan frame a spherically symmetric solution of the type, which is conformally transformed as, 
\be
ds^2=-A(r,t)dt^2+B(r,t)dr^2+r^2d\Omega\ ,
\label{1.6}
\ee
and becomes in the Einstein frame,
\be
ds_E^2=-C(\rho,t')dt'^2+D(\rho,t')d\rho^2+\rho^2d\Omega\ ,
\label{2.8}
\ee
where we have redefined  the coordinates, taking $\rho^2=\Pi^2(r,t)r^2$, and $t'=T(t,r)$ to avoid cross terms, such that the metric can be written in the familiar form as in (\ref{1.6}). It is well known that  the  only vacuum solution for a spherically symmetric metric in General Relativity is given by the Schwarzschild solution, or Schwarzschild-(A)dS solution with a cosmological constant. This result, called Birkhoff's theorem states basically that the metric \eqref{1.6} is time independent in vacuum, $A(r,t)=A(r)$ and $B(r,t)=B(r)$, Ref.~\cite{Birkhoff}. However, the theorem is not valid for actions of the type of \eqref{1.4} (neither for the conformal action) unless strong conditions are firstly assumed (see Ref.~\cite{Faraoni}). Here, we assume that the theory introduces small corrections to General Relativity, so that the zero order solution satisfies the Birkhoff's theorem, and we study the perturbations around.

\section{Birkhoff's theorem in the Einstein and Jordan frames}

Let us start by analyzing the metric (\ref{2.8}) in the Einstein frame and explore the range where  is static. We consider the perturbed metric,
\be
g_{E\mu\nu}=g_{E\mu\nu}^{(0)}+g_{E\mu\nu}^{(1)}\ , \quad \varphi=\varphi^{(0)}+\varphi^{(1)}\ ,
\label{3.1}
\ee
where the set  $\{g_{E\mu\nu}^{(0)}, \varphi^{(0)}\}$ refers to the zero-order solution, while the perturbations are represented by $\{g_{E\mu\nu}^{(1)}, \varphi^{(1)}\}$. The components of the metric can be written as
\be
\left\{\begin{array}{ll}
g_{Et't'}=-C(\rho,t')\simeq -C^{(0)}(\rho,t')-C^{(1)}(\rho,t') \\
g_{E\rho\rho}=D(\rho,t')\simeq D^{(0)}(\rho,t')+D^{(1)}(\rho,t') \\
g_{E\theta\theta}=\rho^2 \\
g_{E\psi\psi}=\rho^2 \sin^2\theta
\end{array}\right.
\label{3.2}
\ee
Hence,  the field equations (\ref{2.4},\ref{2.5}) can be split  into  different orders of perturbations. As we are interested in vacuum solutions,  $T_{E\mu\nu}^{(m)}=0$, and we assume  a background solution given by a constant scalar field $\varphi^{(0)}(\rho,t')=\varphi_0$, the equations (\ref{2.4}, \ref{2.5}) at zero-order are given by,
\be
R_{E\mu\nu}^{(0)}-\frac{1}{2}g_{E\mu\nu}^{(0)}R_E^{(0)}+g_{E\mu\nu}^{(0)} \Lambda=0\ , \quad \frac{dU(\varphi^{(0)})}{d\varphi}=0\ ,
\label{3.8}
\ee
where the cosmological constant is defined as $\Lambda=\frac{1}{2}U_0$. Equations (\ref{3.8}) are exactly the same as Einstein field equations with a cosmological constant, such that the solution is the well known Schwarzschild-(A)dS metric, that represents the zero-order solution for the metric \eqref{3.1},
\be
C^{(0)}(\rho)=[D^{(0)}(\rho)]^{-1}=1-\frac{2\mu}{\rho}-\frac{\Lambda}{3} \rho^2\ ,
\label{3.8a}
\ee
where $\mu$ is an integration constant. Then, at zero-order  the solution satisfies the Birkhoff's theorem, as expected. At first linear order, the equation (\ref{2.4}) is,
\be
R^{(1)}_{E\mu\nu}-\frac{1}{2}U_0g^{(1)}_{E\mu\nu}=0\ .
\label{3.12}
\ee
where we have used the results obtained at zero order (\ref{3.8}), being $U_0'=0$, and $R^{(0)}=2U_0$. The expression (\ref{3.12}) is a linear system of differential equations in $g^{(1)}_{E\mu\nu}$, where the coefficients are given in terms of the zero order  solution (\ref{3.8a}), and whose solutions exhibits a very complex expression. Nevertheless, we can study the system in order to obtain the time dependence of the solution. From the $rt$-equation,
\be
R_{tr}^{(1)}=\frac{1}{r}\frac{\dot{D}^{(1)}}{D^{(0)}}=0\, \quad \rightarrow \quad  g^{(1)}_{E\rho\rho}=D^{(1)}(\rho,t')=D^{(1)}(\rho)\ .
\label{3.12a}
\ee
Hence, the $rr$-component of the metric is time independent. By deriving the $\theta\theta$-equation respect the time, it gives,
\be
\frac{dR^{(1)}_{\theta\theta}}{dt}=-\frac{r}{2D^{(0)}}\frac{d^2}{drdt}\left(\frac{C^{(1)}}{C^{(0)}}\right)=0\ , \quad \rightarrow C^{(1)}(\rho,t')=C^{(0)}(\rho)\left(\alpha(t')+\chi(\rho)\right)\ ,
\label{3.12b}
\ee
It is straightforward to show by  $rr$- and  $tt$-equations that $\alpha(t')$ is an undetermined function of the time $t'$, so that can be taken to be a constant $\alpha(t')=\alpha$, while the functions $D^{(1)}(\rho)$ and $\chi(\rho)$ are solutions of the system of differential equations  (\ref{3.12}). As the solutions (\ref{3.12a}-\ref{3.12b})   are time-independent,  the Birkhoff's theorem holds, and the metric remains static in vacuum for a general scalar-tensor theory given by the action (\ref{2.1}) in the Einstein frame. 
Let us now transform the metric components (\ref{3.12a}-\ref{3.12b}) to the Jordan frame, and check how the metric is affected. In order to obtain the conformal transformation (\ref{2.1}), we have to solve the scalar field equation at first linear order,  $\Box\varphi^{(1)}=U_0'' \varphi^{(1)}$, which can be expanded as,
\be
C^{(0)}(\rho)\frac{d^2\varphi^{(1)}}{d\rho^2}-\left(C'^{(0)}+\frac{2}{r}C^{(0)}\right)\frac{d\varphi^{(1)}}{d\rho}-\frac{1}{C^{(0)}(\rho)}\frac{d^2\varphi^{(1)}}{dt^2}-U_0''\varphi^{(1)}=0\ .
\label{3.14a}
\ee
This equation clearly has a solution of the type  $\varphi^{(1)}(\rho,t')=\varphi^{(1)}_{\rho}\varphi^{(1)}_{t'}$
Hence, the corresponding time part of the equation is given by,
\be
\frac{\ddot{\varphi}^{(1)}_{t'}}{\varphi^{(1)}_{t'}}=k\, \quad \rightarrow \quad \varphi^{(1)}_{t'}=C_1\e^{\sqrt{k} t'}+ C_2\e^{-\sqrt{k} t'}\ ,
\label{3.14c}
\ee
where $C_{1,2}$ and $k$ are constants. 
While the radial dependence of  $\varphi^{(1)}_{\rho}$ can be obtained  by solving the differential equation \eqref{3.14a}.
Hence, the conformal transformation yields,
\be
\Pi^2=\phi=\phi^{(0)}+\phi^{(1)}=\e^{\varphi/\sqrt{3+2\omega}}\simeq \phi^{(0)} \left(1+\frac{1}{\sqrt{3+2\omega}}\varphi^{(1)}(t',\rho)\right)\ ,
\label{3.16}
\ee
where $\phi^{(0)}=\e^{\varphi^{(0)}/\sqrt{3+2\omega}}=constant$. Hence, the conformal transformation is time dependent, and the  metric in the Jordan frame yields,
\be
ds^2=\frac{ds^2_E}{\phi(\rho,t')}\simeq\frac{\phi^{(0)}-\phi^{(1)}(\rho,t')}{\left(\phi^{(0)}\right)^2}\left(-C(\rho)dt'^2+D(\rho)d\rho^2+\rho^2d\Omega\right)\ .
\label{3.17}
\ee
Hence, the perturbations on the metric are not  static at first linear order in the Jordan frame, and the Birkhoff's theorem is not  valid, which differs  from the results in the Einstein frame. Note also that the result given in (\ref{3.14c}) suggests that the zero-order solution will be unstable in the Jordan frame due to the perturbations induced by the scalar field. This fact suggests the non-physical equivalence between both frames, as has already pointed out in Ref.~\cite{JordanVsEinstein}. At zero order, where $\phi^{(0)}$ in (\ref{3.17}) is a constant, the result in the Jordan frame gives also a Schwarzschild-(A)dS metric, and the Birkhoff's theorem is satisfied. Nevertheless, at  first linear order the metric is obviously not static, and the theorem is not satisfied, which contradicts the result obtained in the Einstein frame, and points to the different physical meaning of both frames at least in a perturbative approach. 
\section{Acknowledgments}
We thank professor Sergei Odintsov for discussions and suggestions on the topic. DSG acknowledges a FPI fellowship from MICINN (Spain), project FIS2006-02842.
\bibliographystyle{aipproc}   

\end{document}